\documentclass{article}
\newcommand{\bfr}{\begin{flushright}}
\newcommand{\efr}{\end{flushright}}
 
\begin{document}
\title{Cosmological String Theory with Thermal Energy
}
\author{Kiyoshi Shiraishi\\
Institute for Nuclear Study, University of Tokyo,\\ Midori-cho, Tanashi,
Tokyo 188, Japan
}
\date{Europhysics Letters {\bf 8}  (1989) pp. 303--307
}
\maketitle
\begin{abstract}
An attempt to construct a cosmological scenario directly from string
theory is made. Cosmological string theory was presented by Antoniadis,
Bachas, Ellis and Nanopoulos. They also expect loop effects on
cosmological string theory. In this paper, we point out the other
importance of the one-loop effect, the finite temperature effect. The
equations of motion for background geometry at finite temperature are
given. We address a problem on derivation of the effective action at
nonzero temperature.
\end{abstract}

\bigskip

Nowadays, string theory \cite{1} is seriously considered as a 
candidate for a ``theory of everything''. Such a ``theory'' has to
explain the way of unification of forces as well as the origin of
the universe.

We suspect that some ``stringy'' features must be of importance in
the early stage of the universe, and may solve the problems on inflation
and on the initial singularity.

One can derive a set of
Einstein equations from string theory \cite{2} as an effective field
theory: even if we use the equations with the other aspects of string
theory, we have yet no fully satisfactory 
superstring cosmology \cite{3} however.

In a recent paper, Antoniadis et al. suggested an attempt to consider
the interplay between string theory and cosmology by assuming evolution
equations of the type of Robertson-Walker 
universe in the effective field theory \cite{4}. 

The authors of ref.~\cite{4} insisted on the importance of loop effects
to solve the problem of dilaton potential \cite{5} wich concerns some
aspects of cosmology. If it is the case, why does not the thermal
effect which arises from just the loop effect \cite{6} have some
importance on cosmology? Particularly, in string theory, it is
interesting to take into account the thermal matter because of its
peculiar behaviour at high temperature \cite{7}. It is remarkable that
the authors of ref.~\cite{4} brought various possibilities to
cosmology. It is quite reasonable to respect the finite-temperature
effect on cosmological solutions obtained from their scenarios. 

In the rest of this letter, we introduce temperature into the
cosmological equations in the manner of ref.~\cite{8}. Here, we
consider closed bosonic strings only, though they contain a tachyon,
for a pedagogical reason. To apply the formulation to superstrings is
straightforward. We investigate derivation of a set of Einstein
equations when the background space-time is nearly flat, to make a
first step to the ``stringy'' cosmology.

In string theory equations of motion for background fields
are derived by requiring conformal invariance on sigma models \cite{2}.
As stated in Ref.~\cite{4}, the effective action for these equations
has a relation to the center of Virasoro algebra
\cite{2}. Practically, when we want to get loop corrections to
equations of motion, we apply the following equation to an
appropriate loop order; \cite{9,10}
\begin{equation}
\sum_{genus}\langle V(0)\rangle =0\,.
\label{eq2}
\end{equation}
This must hold for all massless vertex operator $V$;
because of momentum conservation, the vertex operator has
momentum here.

We get along with this line; we follow particularly the
scheme in ref.~\cite{10}. In addition, we consider the graviton
vertex. The use of the dilaton vertex raises problems, which are
related to ones pointed out previously; the connection to the
center of the Virasoro algebra is problematical especially in our
case, as we will mention.

We start with the physical graviton vertex in eq.~(\ref{eq2}). We
treat it at the one-loop level, that is, genus one.

The action for the closed bosonic string in twenty-six
dimensions is given by
\begin{equation}
S=\frac{T}{2}\int d^2z\sqrt{g} g^{ab}\partial_aX^\mu\partial_bX^\nu
G_{\mu\nu}(X)\,,
\end{equation}
where $G$ is the background metric and $T$ is the string tension.
The vertex operator for gravitons is
\begin{equation}
V(p)=T\int d^2z\sqrt{g} g^{ab}\partial_aX^\mu\partial_bX^\nu
\zeta_{\mu\nu}(p)e^{ip\cdot X}\,,
\end{equation}
where the polarization tensor $\zeta$ satisfies $\zeta^{\mu\nu}p_\nu=0$
and
$\zeta^\mu_\mu=0$.

Then the tadpole amplitude can be calculated by using
Polyakov's path integral method (see ref.~\cite{8,11}).  For genus
zero, the world sheet has $S^2$ topology, and then we get
\begin{equation}
\langle V(0; G^0_{\mu\nu})\rangle_{S^2}=0\quad\mbox{for}\quad
G^0_{\mu\nu}=\eta_{\mu\nu}\,,
\end{equation}
where $\eta_{\mu\nu}= diag.(-1,1,\dots,1)$. However, if the space-time
is slightly curved, it yields a non-trivial result,
\begin{equation}
\langle V(0; G_{\mu\nu})\rangle_{S^2}\ne 0\,.
\end{equation}
The deviation $\Delta G_{\mu\nu}=G_{\mu\nu}-\zeta_{\mu\nu}$ is
presumably of order of the loop coupling constant $g^2$. The above is
compensated by the one-loop, one-partle amplitude
\begin{equation}
\langle V(0; G^0_{\mu\nu})\rangle_{T^2}\ne 0\,,
\end{equation}
where $T^2$ denotes two-torus. Therefore we
observe the following
equation to the one-loop level:
\begin{equation}
\langle V(0; G_{\mu\nu})\rangle_{S^2}+\langle V(0;
G^0_{\mu\nu})\rangle_{T^2}=0\,.
\end{equation}

In the curved space-time, the
fact that $\langle V(0; G_{\mu\nu})\rangle_{S^2}$ becomes
finite comes from the cancellation of logarithmic divergences from
the volume of M\"obius transformations and the short-distance limit of
the  two-point  correlation  function \cite{10}.  After regularization,
we can get \cite{9,10}
\begin{equation}
\langle V(0; G_{\mu\nu})\rangle_{S^2}=cg^{-2}\zeta^{\mu\nu}R_{\mu\nu}\,,
\label{eq9}
\end{equation}
where $R_{\mu\nu}$ is Ricci tensor constucted from $G_{\mu\nu}$ and $c$
is a calculable constant under a certain regularization scheme.

On the other hand, we can incorporate temperature into the
calculation of the one-loop tadpole amplitude $\langle V(0;
G^0_{\mu\nu})\rangle_{T^2}$, by 
making use of the imaginary-time method in field theory \cite{6,8}.

The amplitude at zero temperature is formally written as
follows \cite{8}:
\begin{eqnarray}
\langle V(0)\rangle&=&\int d^2\tau (\det{}' P^+P)^{1/2}\left(
\frac{2\pi}{\int d^2z}\det{}'\Delta\right)^{-13}\nonumber\\
& &\cdot\frac{1}{\Omega}\int DX^\mu\, e^{-S} \,\zeta_{\mu\nu}\, \int
d^2z
\,T\langle\partial_aX^\mu\partial^aX^\nu\rangle\,,
\end{eqnarray}
where $\Omega$ is the gauge volume; for other notations, see
\cite{8,11}. Actually, this diverges because of the existence of a
tachyon.

The prescription given by Polchinski \cite{8} is that the zero-mode
of the ``time-coordinate'' possesses a periodic structure in terms
of the two dimensional coordinates $z^2$, i.e.,
\begin{equation}
X^0(z^1,z^2+1)=X^0(z^1,z^2)+r\beta\,,
\end{equation}
where $r$ is an integer (winding number) and $\beta$ is regarded as the
inverse of temperature. At the same time, we must set $\eta^{00}=+1$
to take the Euclidean signature. Accordingly, the zero-mode part
emerges in the action \cite{8}, and furthermore, the correlation
function of the ``time'' component is regarded as:
\begin{eqnarray}
\langle\partial_aX^\mu\partial^aX^\nu\rangle&=&
+\frac{\eta^{\mu\nu}}{T\tau_2}+\frac{r^2\beta^2}{\tau_2^2}
\delta^{\mu 0}\delta^{\nu 0}\nonumber \\& &+
\mbox{(the divergence from a delta-function)}\,.
\end{eqnarray}

Owing to this, the result of the calculation of the amplitude at finite
temperature turns out to be Lorentz noncovariant
\begin{equation}
\langle V(0)\rangle_{T^2}=-\zeta^{00}T_{00}-\zeta^{ij}T_{ij}\,,
\label{eq13}
\end{equation}
where
\begin{eqnarray}
T_{00}&=&T^{13}\int_0^\infty\frac{d\tau_2}{2\pi\tau_2^2}
\int_{-1/2}^{1/2}d\tau_1 e^{4\pi\tau_2}(2\pi\tau_2)^{-12}
\left|\prod_{n=1}^\infty(1-e^{2\pi in\tau})\right|^{-48}\nonumber \\
& &\qquad\cdot\sum_{r=1}^\infty e^{-r^2\beta^2 T/2\tau_2}\left(1+
\frac{r^2\beta^2 T}{\tau_2}\right)\,,
\label{eq14}
\end{eqnarray}
while
\begin{eqnarray}
T_{ij}&=&\delta_{ij}\,T^{13}\int_0^\infty\frac{d\tau_2}{2\pi\tau_2^2}
\int_{-1/2}^{1/2}d\tau_1 e^{4\pi\tau_2}(2\pi\tau_2)^{-12}
\left|\prod_{n=1}^\infty(1-e^{2\pi in\tau})\right|^{-48}\nonumber \\
& &\qquad\cdot\sum_{r=1}^\infty e^{-r^2\beta^2 T/2\tau_2}\,,
\label{eq15}
\end{eqnarray}
where $i, j=1,2,\dots,25$.

In each equation above, the volume factor is discarded; it can be
absorbed into the Newton constant.

Consequently, from eqs.~(\ref{eq9}), (\ref{eq13}), (\ref{eq14}), and
(\ref{eq15}), we can express the effective equation as
\begin{equation}
R_{\mu\nu}=
c'T_{\mu\nu}+{\rm (terms~proportional~to~}\eta_{\mu\nu})\,,
\label{eq16}
\end{equation}
where $c'$ is a constant of order $g^2$. Because of tracelessness of the
physical graviton, we failed to determine the part proportional to the
metric tensor in this level. Of course by demanding the conservation of
energy-momentum, we can reconstruct the full Einstein equation
\begin{equation}
R_{\mu\nu}=\kappa^2\left(T_{\mu\nu}-\frac{1}{D-2}G_{\mu\nu}
T^\lambda_\lambda\right)\,,
\end{equation}
with $D=26$. Here $\kappa^2=8\pi G$; $G$ is the Newton constant. These
are precisely the expected results. $T_{\mu\nu}$ is written by
\begin{equation}
T^\mu_\nu=diag.(-\rho, p, \dots, p)\,,
\end{equation}
where the energy density $\rho$ and the pressure $p$ can be derived
from the free energy density $f$ originally calculated in ref.~\cite{8}:
\begin{eqnarray}
f&=&-T^{13}\int_0^\infty\frac{d\tau_2}{2\pi\tau_2^2}
\int_{-1/2}^{1/2}d\tau_1 e^{4\pi\tau_2}(2\pi\tau_2)^{-12}
\left|\prod_{n=1}^\infty(1-e^{2\pi in\tau})\right|^{-48}\nonumber \\
& &\quad\cdot\sum_{r=1}^\infty e^{-r^2\beta^2 T/2\tau_2}\,,
\label{eq19}
\end{eqnarray}
by the relation in thermodynamics
\begin{equation}
\rho=f-(\beta^{-1})\frac{\partial f}{\partial(\beta^{-1})}\quad
\mbox{and}\quad
p=-f\,.
\end{equation}

Now we must make remarks: how can we derive the
effective action for background fields directly  from  the
calculation of dilaton tadpole amplitudes, which is expected to
be connected with the central charge of Virasoro algebra? That
amplitude might contain the part, which could not be derived above and
is proportional to the metric tensor of the Einstein equation. However,
at finite temperature, there are two possibilities: the
amplitude includes whether the trace of the energy-momentum tensor or
the free energy.

The origin of the problem must be attributed partially to the lack of
the information of background fields in vertex functions. In the level
of approximation here, we used the vertex functions in flat space-time;
in the case treated above, the polarization tensor $\zeta$ is
constrained to obey $\mbox{tr}\,\zeta=0$ and $p\cdot\zeta=0$. This
cannot play the full role of the external sources in partition
functions. The same is true for the dilaton vertex and the situation is
rather worse due the ambiguity in its form even in flat space-time.
Moreover, the construction of the vertex operators in curved spaces is
not sufficiently understood at least for the present perpose.

Therefore, to my knowledge, it is difficult to give the  answer
definitively. It might be guessed that a more suitable
form of the vertex operator would exist. This problem is related to
the problem referred in ref.~\cite{12}.

The derivation of Einstein equations above is mathematically
well defined except for the actual divergence due to a tachyon. But
this has no meaning in physics. As one might have already been
aware, there are contributions from gravitons and from infinitely
massive states in the right-hand side of eq.~(\ref{eq16}).
In our universe, the contributions must have departed from
equilibrium; moreover the inclusion of gravity in thermal physics
brings about difficulties \cite{13}. Thus we must seriously consider
the non-equilibrium process and the dynamics in the universe in order
to make the equations physically meaningful.

In addition, the Einstein equations derived here are very trivial ones.
To take genuine ``stringy'' effects into consideration, of
course, we have to investigate all loop order or nonperturbative
effects. Nevertheless the possibility of occurrence of inflation
pointed out in ref.~\cite{4} seems to be an interesting suggestion.
In this case the inclusion of temperature is necessary for constructing
a more or less realistic stringy cosmology.


\section*{Acknowledgments}
This work is supported in part by the Grant-in-Aid for Encouragement
of Young Scientist from the Ministry of Education, Science and Culture
(\# 63790150).

The author is grateful to the Japan Society for the Promotion
of Science for the fellowship. He also thanks Iwanami F\=ujukai
for financial aid.


\end{document}